\documentclass[12pt]{article}

\usepackage{amsmath,amsfonts,amssymb,graphicx,psfrag,rotating}
\usepackage[colorlinks=true,urlcolor=blue,citecolor=black,linkcolor=blue]{hyperref}
\usepackage{subfigure}
\usepackage{setspace} 
\usepackage{dcolumn}

\usepackage{natbib}
\bibpunct{(}{)}{;}{a}{}{,}

\usepackage{comment}
\usepackage{multirow}
\usepackage{lscape}
\usepackage{bbm}
\usepackage{scalefnt}
\usepackage[autolanguage]{numprint}
\usepackage{caption}
\usepackage{booktabs}
\usepackage{bm}
\usepackage{threeparttable}
\usepackage{rotating}
\usepackage{color}
\usepackage{setspace}

\usepackage{lmodern}
\fontfamily{lmtt}\selectfont
\usepackage[T1]{fontenc}

\bibpunct{(}{)}{;}{author-year}{}{,}

\usepackage[top=1in, left=1in, right=1in,
bottom=1in]{geometry}

\makeatletter
\renewcommand{\section}{\@startsection{section}{1}{0em}{\baselineskip}{0.5\baselineskip}{\large\bfseries\large}}
\renewcommand{\subsection}{\@startsection{subsection}{0}{0em}{\baselineskip}{0.5\baselineskip}{\normalfont\bfseries\normalsize}}
\makeatletter

\newcolumntype{.}{D{.}{.}{-1}}
\newcolumntype{d}[1]{D{.}{.}{#1}}

\begin{document}
\pagestyle{plain}

\newcommand{\blind}{0}

\newcommand{\tit}{\large{Inference for Instrumental Variables: A Randomization Inference Approach}}

\if0\blind

{\title{\tit\thanks{We thank Frank Windemeijer, Guido Imbens, Jacob Klerman and participants of the University of Pennsylvania Causal Inference Group for comments and suggestions. The research of Hyunseung Kang was supported in part by NSF Grant DMS-1502437. The research of Luke Keele was supported in part by U.S. Department of Labor Grant GS-10F-0086K.}}
\author{Hyunseung Kang\thanks{Assistant Professor, University of Wisconsin, Madison, Email: hyunseung@stat.wisc.edu}
\and Laura Peck\thanks{Principal Scientist, Abt Associates, Washington D.C., Email: Laura\_Peck@abtassoc.com}
\and Luke Keele\thanks{Professor, Georgetown University, Email:
      lk681@georgetown.edu}}

\date{\today}

\maketitle
}\fi

\if1\blind
\title{\bf \tit}
\maketitle
\fi

\maketitle

\thispagestyle{empty}

\begin{abstract}
The method of instrumental variables (IV) provides a framework to study causal effects in both randomized experiments with noncompliance and in observational studies where natural circumstances produce as-if random nudges to accept treatment. Traditionally, inference for IV relied on asymptotic approximations of the distribution of the Wald estimator or two-stage least squares, often with structural modeling assumptions and/or moment conditions. In this paper, we utilize the randomization inference approach to IV inference. First, we outline the exact method, which uses the randomized assignment of treatment in experiments as a basis for inference, but lacks a closed-form solution and may be computationally infeasible in many applications. We then provide an alternative to the exact method, the almost exact method, which is computationally feasible but retains the advantages of the exact method. We also review asymptotic methods of inference, including those associated with two-stage least squares, and analytically compare them to randomization inference methods. We also perform additional comparisons using a set of simulations. We conclude with three different applications from the social sciences. 
\end{abstract}

\doublespacing

\section{Introduction} \label{sec:intro}

\subsection{Instrumental Variables: A General Causal Method} 
\label{sec:introMotivation}

Many randomized trials suffer from noncompliance where subjects fail to comply with his or her assigned treatment status. While analysts can focus on the causal effect of the treatment assignment on the outcome in an intention-to-treat (ITT) analysis, there is also substantial interest in the causal effect of the treatment actually received, especially in the social sciences and public policy/program evaluations. For example, as part of a comprehensive economic stimulus package funded under the 2009 American Recovery and Reinvestment Act (ARRA), the U.S. Department of Labor awarded a series of grants across the United States to promote training for employment in energy efficiency, renewable energy, and health care and participants were randomized to either participation in the new training programs, i.e. treatment, or to the existing training programs available---control; see Section \ref{sec:app1} for details. However, some trainees who were assigned to the new training initiatives selected not to participate, creating issues of non-compliance, and the Green Jobs and Health Care (GJ-HC) Impact Evaluation \citep{Copson:2015,Martinson:2015} wanted to evaluate the actual efficacy of the training programs, i.e. whether those who actually received the training programs had any impact on labor outcomes like earnings and employment outcomes. 


When noncompliance is present, substantial progress can be made by using the treatment assignment as an instrumental variable (IV), which is a variable that affects exposure to treatment but does not directly affect the outcome \citep{Angrist:1996,Hernan:2006}; see Section \ref{sec:IVAssumption} for details about IV and related assumptions. These identifying assumptions allow one to estimate the complier average causal effect (CACE), which is the average causal effect among subpopulation of individuals who comply with the treatment assignment, or the average treatment effect among the treated (ATT). However, IV analysis is not confined to randomized trials with non-compliance. In some applications, the instrument is some naturally occurring nudge to accept a treatment, and is characterized as a type of natural experiment. In observational studies, IV analysis may provide additional insights about the causal effects in the presence of unobserved confounding \citep{kang_commentary_2016}. As such, IV analysis is utilized in many disciplines including economics \citep{Angrist:2001,Imbens:2014}, epidemiology \citep{Hernan:2006, Baiocchi:2014}, political science \citep{Hansford:2010,Keele:2014b}, and Mendelian randomization (MR) studies \citep{davey_smith_mendelian_2003, davey_smith_mendelian_2004, lawlor_mendelian_2008} where the instruments are genetic variants. 
 
\subsection{Traditional Approaches to Inference with Instrumental Variables}
 
Once identifying assumptions are accepted, point estimation of the casual effect is typically based on the Wald estimator \citep{Wald:1940}. In the case of randomized experiments with a binary treatment assigned indicator/IV, the Wald estimator is equivalent to the popular two-stage least squares (TSLS) estimator in the IV literature \citep{Angrist:2008}; see Section \ref{sec:wald} for details. For inference, such as confidence intervals, traditional methods rely on econometric theory by using structural models, moment assumptions, and/or asymptotic approximations \citep{Angrist:2008,Wooldridge:2010}. For example, under the three moment restrictions on the error terms in a structural model, specifically 2SLS.1, 2SLS.2, and 2SLS.3 in Chapter 5 of \citet{Wooldridge:2010}, the TSLS estimator is asymptotically normal. The asymptotic normality can be used to derive inferential quantities like p-values and confidence intervals and is the default mode of inference in most statistical software. 

Despite the simplicity and familiarity of the Wald approach to inference, especially given its ties to standard asymptotically Normal-based testing procedures, inference for IV methods is complicated by the fact that statistical uncertainty depends not only on the sample size, but also on the magnitude of the effect of the instrument on the treatment, commonly known as the strength of an instrument. Generally speaking, when the instrument has a large effect on the treatment, it is known as a strong instrument, while a small effect is known as a weak instrument. It is well known in the literature, that even in large ``asymptotic'' samples, confidence intervals based on TSLS will have incorrect coverage when the instrument is weak \citep{nelson_some_1990,Bound:1995,Staiger:1997,Dufour:1997} and some proposals have been suggested to remedy this problem \citep{anderson_estimation_1949,zivot_valid_1998,wang_inference_1998, kleibergen_pivotal_2002, moreira_conditional_2003}. Unfortunately, these results are derived from structural models, moment assumptions, and/or asymptotic approximations. For example, the conditional likelihood ratio approach proposed by \citet{moreira_conditional_2003} relies on a specification of a structural model as well as Normal structural errors (for finite-sample behavior) or asymptotic moment conditions on the errors (for asymptotic behavior). Most recently, in Mendelian randomization where weak instruments are common, several proposals have been suggested \citep{burgess_avoiding_2011, burgess_bias_2011, pierce_power_2011} to remedy the bias induced by weak instruments, most notably by finding additional instruments, combining them to a single score/instrument, and using the Wald approach to inference. \citet{burgess_improving_2012} also suggest using a Bayesian approach to deal with weak instruments.

\subsection{Our Contributions}

Given the complexities of inference with IV methods, this paper reviews and extends an alternative, unified framework for inference that is based solely on the assumed instrument assignment mechanism. Specifically, we discuss exact and almost exact inferential methods for IV. The exact method uses randomization as the ``reasoned basis for inference'' \citep{Fisher:1935} and mirrors the original design of a randomized experiment. The exact method produces an honest confidence interval for the target causal parameter even when the causal effect of the instrument on the treatment is weak \citep{Imbens:2005b,Keele:2016a}. Also, unlike the aforementioned standard methods based on large sample Normal approximations which assume that the participants are a random sample from the target population, the exact method is finite-sample based and makes it explicit that further assumptions are required to generalize the IV estimand to other populations. 

The second method, the almost exact method, behaves like the exact method, except that it avoids the computationally intensive nature of the exact method in larger sample sizes. The almost exact method is also motivated by the design of the instrument assignment mechanism and is based on finite sample asymptotics \citep{Hajek:1960}. By doing so, the almost exact method preserves the properties of the exact method, but is computationally tractable with a closed-form expression. We also discuss extensions to both methods, including adjustment of pre-treatment covariates, choice of test statistics, multiple treatments, and sensitivity analysis. Importantly, we do not rely on the assumption of constant treatment effect, which is typical in randomization-based inference; see Sections \ref{sec:CIFisher} and \ref{sec:AlmostExact}. 

Next, we analytically and numerically compare these randomization-based methods with traditional methods including the TSLS estimator. We show how popular inferential methods in instrumental variables can be derived using our framework and are special cases of our randomization inference approach. We also show that the traditional approaches are, in essence, approximations of the exact method with varying degrees of accuracy and complexity. We, then, highlight the strengths of the exact and almost exact approach in three empirical applications. Two of the applications are based on randomized trials with noncompliance, and the third is an observational study based on a natural experiment. Finally, we include \texttt{R} code online for others to use.

The comparison of exact and almost exact methods to traditional approaches to inference reveals the strengths of the randomization-inference approach in these settings, which are (1) the basic assumptions of inference are transparent and based on the study design, not on structural assumptions and assumptions about moments of structural error terms, (2) these methods always provide honest inference in the form of correct Type I error control, and (3) they make it clear where traditional assumptions like homoskedastic errors and/or large $n$ play a critical role in about inference for the causal effect.

\section{Framework: Notation, Assumptions, and Estimators}
\subsection{Notation} 
\label{sec:notation}

Suppose there are $n$ individuals in an experiment indexed by $i = 1,\ldots,n$. Let $Y_i$ denote the outcome, $D_i$ denote the treatment (or treatment received), and $Z_i$ denote a binary instrument (or treatment assignment indicator). For each individual $i$, we observe the triplet $(Y_i, D_i, Z_i)$. Also, for each individual $i$, let $Y_i^{(z,d)}$ be the potential outcome of the outcome given the instrument value $z \in \{0,1\}$ and treatment value $d \in \{0,1\}$ and let $D_i^{(z)}$ be the potential outcome of the treatment given the instrument value $z \in \{0,1\}$. The relationship between the potential outcomes, $Y_i^{(z,d)}$ and $D_i^{(z)}$, and the observed triplets, $(Y_i, D_i, Z_i)$ is: 
\[
D_i = D_i^{(Z_i)} = Z_i D_{i}^{(1)}  + (1 - Z_i) D_{i}^{(0)}, \quad{} Y_i  = Y_i^{(Z_i,D_i)} =  Y_i^{\left(Z_i, D_i^{(Z_i)}\right)} = Z_i Y_{i}^{\left(1,D_i^{(1)} \right)} + (1 - Z_i) Y_i^{\left(0,D_{i}^{(0)} \right)}
\]
Our notation implicitly assumes the stable unit treatment value assumption (SUTVA) \citep{Rubin:1980}. Let $\mathcal{F}= \{(Y_i^{(1,1)}, Y_{i}^{(1,0)}, Y_{i}^{(0,1)}, Y_i^{(0,0)}, D_{i}^{(1)}, D_{i}^{(0)}),i=1,\ldots,n\}$ denote the collection of potential outcomes for all $n$ individuals. Also, let $0 < n_1 < n$ represent the number of individuals assigned to treatment $Z_{i} = 1$ and $n_0 = n - n_1$ represent the number of individuals assigned to control $Z_{i} = 0$ where $n_1$ and $n_0$ are non-random. Let $\Omega = \{(z_1,\ldots,z_n) \in \{0,1\}^{n}, \sum_{i=1}^{n} z_{i} = n_1\}$ be the possible values that $(Z_1,\ldots,Z_n)$ can take so that among $n$ individuals, exactly $n_1$ individuals have $Z_{i} = 1$ and the rest $n_0 $ individuals have $Z_{i} = 0$. 

Let $\mathcal{Z}$ be the event $(Z_1,\ldots,Z_n) \in \Omega$. Following our discussion about randomization as a basis for inference, the paper will focus on the finite population settings where the target of inference is a functional of $\mathcal{F}$ that is fixed and unknown. Extensions to inference based on infinite population models is possible and is detailed in Chapter 6 of \citet{Imbens:2015}. Such approaches typically require additional assumptions such as the study participants are a random sample from the target population. Here, we wish to be explicit that further assumptions will be required to generalize the causal quantities of interest to other populations.

\subsection{Causal Estimands and Instrumental Variables Assumptions} 
\label{sec:IVAssumption}

Given the potential outcomes in $\mathcal{F}$, we can define the following causal estimands.
\begin{align}
\tau_{Y} &= \frac{1}{n} \sum_{i=1}^{n} Y_i^{(1,D_i^{(1)})} - Y_{i}^{(0,D_i^{(0)})} \label{eq:tau_Y}\\
\tau_{D} &= \frac{1}{n} \sum_{i=1}^{n} D_i^{(1)} - D_{i}^{(0)} \label{eq:tau_D} \\
\tau &= \frac{\tau_{Y}}{\tau_{D}} = \frac{\sum_{i=1}^{n} Y_i^{(1,D_i^{(1)})} - Y_{i}^{(0,D_i^{(0)})}}{ \sum_{i=1}^{n}D_i^{(1)} - D_{i}^{(0)}} \label{eq:tau_n}
\end{align}
Equation \eqref{eq:tau_Y} is the average causal effect of the instrument on the outcome and is often referred to as the intent-to-treat (ITT) effect. Equation \eqref{eq:tau_D} is the average causal effect of the instrument on the exposure. If, in a randomized experiment, the compliance to assigned treatment is one-sided where individuals assigned to control cannot actually receive the treatment, $D_{i}^{(0)} = 0$, $\tau_{D}$ is known as the compliance rate. Equation \eqref{eq:tau_n} is the ratio of the two average causal effects $\tau_{Y}$ and $\tau_{D}$ and is the causal estimand of interest where it is implicitly assumed that $\tau_{D} \neq 0$; see assumptions (A2) and (A4) below. The estimand $\tau$ is also referred to as the IV estimand in the literature.

Identification of $\tau$ requires a set of assumptions. For example, if we assume (A1) ignorability of $Z$ where $P((Z_1,\ldots, Z_n) = (z_1,\ldots,z_n) | \mathcal{F}, \mathcal{Z}) = P((Z_1,\ldots, Z_n) = (z_1,\ldots,z_n) | \mathcal{Z}) = 1/{{n \choose n_1}}$, (A2) a non-zero causal effect of $Z$ on $D$ where $\tau_{D} \neq 0$, (A3) exclusion restriction where for all $d$ and $i$, $Y_{i}^{(d,1)} = Y_{i}^{(d,0)}$, and (A4) monotonicity where for all $i$, $D_{i}^{(1)} \geq D_{i}^{(0)}$, $\tau$ is identified as the complier average treatment effect (CATE), or the average treatment effect among those individuals in the experiment who complied with their treatment assignment \citep{Angrist:1996}. Alternatively, one can make different sets of assumptions to identify a different interpretation of $\tau$. For example, if we assume (A1)-(A3), and assume an additive structural mean model for $Y$, $D$, and $Z$, with no effect modification assumption on $Y$ via interactions between $D$ and $Z$, $\tau$ would be the average treatment effect among treated individuals (ATT) \citep{Hernan:2006}. Alternatively, if we only assume (A1) and (A2), $\tau$ would simply be identified as the ratio of two average treatment effects \citep{Baiocchi:2010, Kang:2015}. 

Assumptions (A1), (A2) and (A4) are typically satisfied by design of many randomized experiments, especially in public policy program evaluations where there is often one-sided compliance. In natural experiments or observational studies, these assumptions require careful consideration. For example, (A1) holds by design in a randomized experiment, but requires justification when IV is applied to observational data. In Section \ref{sec:sens}, we discuss a sensitivity analysis that allows investigators to probe this assumption.

Since the focus of the paper is on inference for $\tau$, we will not dwell on identifying assumptions. We will simply assume that (A1)-(A4) hold so that $\tau$ is identified as CATE and to allow easier comparison between our randomization inference approach and the traditional approaches to inference. But, we stress that these assumptions are actually more stringent than is necessary for inference on $\tau$ under the approach we use. In particular, in Section \ref{sec:ourapproach}, we will show that the exact and the almost exact methods only rely on (A1) and is robust to near-violations of (A2). Inference from both methods remain valid even when (A3) and/or (A4) may not hold. For additional discussions on these assumptions, see the Supplementary Materials, \citet{Angrist:1996}, \citet{Hernan:2006}, \citet{Deaton:2010}, \citet{Imbens:2010}, \citet{Imbens:2014}, \citet{Baiocchi:2014} and \citet{Swanson:2014}. 

\subsection{Point Estimator for $\tau$} \label{sec:wald}
To discuss inference for $\tau$, especially comparing the randomization inference approach to the traditional approaches based on point estimators (e.g. TSLS) in Section \ref{sec:CIAsymptotic}, it is instructive to discuss point estimators for $\tau_{Y}$ and $\tau_{D}$ that make up $\tau$. The most popular point estimators for $\tau_Y$ and $\tau_D$ are the difference in means estimators.
\begin{align}
\hat{\tau}_{Y} &= \frac{1}{n_1} \sum_{i=1}^{n} Y_i Z_i - \frac{1}{n_0} \sum_{i=1}^{n} Y_i (1 - Z_i) \label{eq:tauhat_Y}\\
\hat{\tau}_{D} &= \frac{1}{n_1} \sum_{i=1}^n D_i Z_i - \frac{1}{n_0} \sum_{i=1}^{n} D_i (1 - Z_i) \label{eq:tauhat_D}
\end{align}
Standard arguments can show that $\hat{\tau}_{Y}$ and $\hat{\tau}_{D}$ are unbiased estimators of $\tau_{Y}$ and $\tau_{D}$, respectively, i.e. $E(\hat{\tau}_{Y} | \mathcal{F}, \mathcal{Z}) = \tau_{Y}$ and $E(\hat{\tau}_{D} | \mathcal{F}, \mathcal{Z}) = \tau_{D}$ \citep{Imbens:2015}. Standard estimators for $Var(\hat{\tau}_{Y} | \mathcal{F}, \mathcal{Z})$ and $Var(\hat{\tau}_{D} | \mathcal{F}, \mathcal{Z})$ exist depending on the assumptions one makes about $\mathcal{F}$ \citep{Imbens:2015}. For now, we will leave them unspecified and denote the estimated variances of $\hat{\tau}_{Y}$ and $\hat{\tau}_{D}$ as  $\widehat{Var}(\hat{\tau}_{Y} | \mathcal{F}, \mathcal{Z})$ and $\widehat{Var}(\hat{\tau}_{D} | \mathcal{F}, \mathcal{Z})$, respectively.

Given the estimators in equations \eqref{eq:tauhat_Y} and \eqref{eq:tauhat_D} for $\tau_{Y}$ and $\tau_{D}$, respectively, the most natural estimator for $\tau$ would be the ratio of the estimators. Indeed, this is the most frequently used estimator for $\tau$ and is often called the ``usual'' IV estimator, ``the'' IV estimator, or the ``Wald'' estimator \citep{Wald:1940, Hernan:2006, Wooldridge:2010, Baiocchi:2014}
\begin{equation} \label{eq:tauhat_n}
\hat{\tau} = \frac{\hat{\tau}_{Y}}{\hat{\tau}_{D}} = \frac{\frac{1}{n_1} \sum_{i=1}^{n} Y_i Z_i - \frac{1}{n_0} \sum_{i=1}^{n} Y_i (1 - Z_i)}{ \frac{1}{n_1} \sum_{i=1}^n D_i Z_i - \frac{1}{n_0} \sum_{i=1}^{n} D_i (1 - Z_i) }
\end{equation}
One can also arrive at $\hat{\tau}$ by using TSLS, another popular point estimator for $\tau$ \citep{Wooldridge:2010, Baiocchi:2014}. Specifically, if one (i) fits a linear regression between $Z_i$ and $D_i$ and saves the predicted $D_i$s, and (ii) fit a second linear regression between the predicted $D_i$ and $Y_i$, the coefficient associated with the predicted $D_i$ from the second linear regression is $\hat{\tau}$.

\section{Inference for $\tau$ Using the Randomization-Based Approach} \label{sec:ourapproach}
In this section, we summarize two methods of inference for $\tau$, both motivated by randomization-based inference. The first method, which we call the exact approach, is guaranteed to have correct coverage. Unfortunately, it is computationally intensive for even modest sample sizes and lacks a closed-form expression. We then outline an approximation to the exact method which we call the almost exact method. In both cases, we only use the design assumptions, primarily (A1), to derive inferential quantities like confidence intervals and p-values.

\subsection{The Exact Method} 
\label{sec:CIFisher}
One method of inference for $\tau$ is based on randomization-based inference approach to instrumental variables as described in \citet{Rosenbaum:1996}, \citet{Rosenbaum:2002}, \citet{Imbens:2005b}, \citet{Baiocchi:2010}, \citet{nolen2011randomization} and \citet{Kang:2015}. It is also called the ``exact'' method because inference relies on exact calculations based on the distribution of $Z$. As we outline below, the randomization inference test of no effect can be inverted to provide distribution-free confidence intervals. 

Typically, under the exact approach, inverting exact tests to derive confidence intervals is associated with the assumption that the treatment effect is constant from unit to unit out of convenience. This assumption of constant treatment effects is often critiqued as unrealistic given the likely presence of ``essential heterogeneity,'' where individuals who will benefit more from a given treatment are more likely to actually seek treatment \citep{heckman2006understanding}. An expanding literature demonstrates how exact inference is possible without constant effect assumptions; see \citet{Rosenbaum:2001},\citet{Rosenbaum:2003}, \citet{Keele:2016a} and \citet{ding2015randomization} for examples. In what follows, we do not assume constant treatment effects.
  
Formally, given $\mathcal{F}$ and $\mathcal{Z}$, consider the null hypothesis $H_0: \tau = \tau_{0}$ which imposes structure on $\mathcal{F}$. This is a composite null hypothesis because there are several values of $\mathcal{F}$ for which the null can be true. Also, $H_0$ is not a sharp null hypothesis \citep{Fisher:1935} whereby a sharp null would allow us to infer other values of the unobserved potential outcomes. In fact, the sharp null of no ITT effect, $Y_{i}^{(1,D_{i}^{(1)})} = Y_{i}^{(0,D_{i}^{(0)})}$ for all $i$ implies $H_0:\tau = 0$, but the converse is not necessarily true; there can be other values of $\mathcal{F}$ that satisfies the null hypothesis $H_0: \tau = 0$.

Given $H_0$, consider the test statistic $T(\tau_0)$ of the form
\begin{equation}\label{eq:testStat}
T(\tau_0) = \frac{1}{n_1} \sum_{i=1}^{n}Z_i(Y_i - D_i \tau_0) - \frac{1}{n_0} \sum_{i=1}^{n} (1 - Z_i) (Y_i - D_i \tau_0)
\end{equation}
Let $Q_i(\tau_0) =  (Y_i - D_i \tau_0)$, $\bar{Q}^{(1)}(\tau_0) = 1/n_1 \sum_{i=1}^{n}Z_i(Y_i - D_i \tau_0)$, and $\bar{Q}^{(0)}(\tau_0) = 1/n_1 \sum_{i=1}^{n}(1 - Z_i)(Y_i - D_i \tau_0)$. \citet{Rosenbaum:2002} calls  $Q_i(\tau_0)$ an adjusted response where the outcome, $Y_i$, is adjusted by the treatment actually received, $D_i$, based on the value of the null $\tau_0$, i.e. $Y_i - D_i \tau_0$. Then, $\bar{Q}^{(1)}(\tau_0)$ represents the sample average of the adjusted responses $Q_{i}(\tau_0)$ for individuals who were assigned treatment $Z_{i} = 1$ and $\bar{Q}^{(0)}(\tau_0)$ represents the sample average of the adjusted responses for individuals who were assigned control $Z_{i} = 0$. We can also rewrite the test statistic in \eqref{eq:testStat} as the difference between the sample averages of the adjusted responses, i.e. $T(\tau_0) = \bar{Q}^{(1)}(\tau_0) - \bar{Q}^{(0)}(\tau_0)$.

Also consider an estimator for the variance of the test statistic $T(\tau_0)$, denoted as $S^2(\tau_0)$
\begin{equation} \label{eq:varTestStat}
S^2(\tau_0) = \frac{1}{n_1(n_1 - 1)} \sum_{i=1}^{n}Z_i \left(Q_i(\tau_0)- \bar{Q}^{(1)}(\tau_0)\right)^2 + \frac{1}{n_0(n_0 -1)} \sum_{i=1}^{n} (1 - Z_i) \left(Q_i(\tau_0) - \bar{Q}^{(0)}(\tau_0)\right)^2
\end{equation}
In the Supplementary Materials, we show that under the null hypothesis, the test statistic $T(\tau_0)$ is zero and hence, any deviation of $T(\tau_0)$ away from zero, positive or negative, suggests $H_0$ is not true. This observation leads us to reject the null if 
\begin{equation} \label{eq:testStatNull}
P_{H_0}\left(\left| \frac{T(\tau_0)}{S(\tau_0)}\right| \geq t | \mathcal{F}, \mathcal{Z}\right)
\end{equation}
 is less than some pre-specified threshold $\alpha$; for simplicity, we assume the rejection region is symmetric around zero under $H_0$. Here, $t$ in equation \eqref{eq:testStatNull} is the observed value of the standardized deviate $T(\tau_0)/S(\tau_0)$ and the probability distribution is under the null hypothesis. Also, one can use the duality between testing and confidence intervals to obtain a confidence intervals for $\tau$ \citep{Lehmann:2006,Lehmann:2008}. Specifically, the exact $1 - \alpha$ confidence interval for $\tau$ would be the set of values $\tau_0$ where
\begin{equation} \label{eq:CIExact}
\left\{\tau_0 : P_{H_0}\left(\left| \frac{T(\tau_0)}{S(\tau_0)}\right| \leq q_{1-\alpha/2} | \mathcal{F}, \mathcal{Z}\right)  \right\}
\end{equation}
where $q_{1-\alpha/2}$ is the $1-\alpha/2$ quantile of the null distribution of $T(\tau_0)/S(\tau_0)$. The confidence interval in equation \eqref{eq:CIExact} is exact, as in it only uses the distribution of $\mathbf{Z}$ as outlined in assumption (A1) and makes no additional distributional assumptions. Importantly, it does not assume the exclusion restriction (A3) although the interpretation of $\tau$ would change if we assume (A3); see Section \ref{sec:IVAssumption} for details. The interval is also is honest, as in equation \eqref{eq:CIExact} will cover the true $\tau$ with at least $1 -\alpha$ probability in finite sample. 

There are additional advantages of the inference method outlined above. First, the confidence interval for $\tau$ may be either empty in length \citep[ch. 4]{Rosenbaum:2002}. The confidence interval may be empty if the adjustment for outcomes is far wrong. This could happen if the instrument strongly predicts the outcome but the treatment dosage does not. Second, the confidence interval may be infinite in length if the instrument is weak. Formally, a weak instrument is an instrument $Z$ that is weakly related to $D$ so that $\tau_D$ is close to $0$ and is a near violation of (A2) \citep{Imbens:2005b}. In other words, an instrument is weak when most units ignore the encouragement to take the treatment. Under randomization inference, if the instrument is weak, the interval becomes longer and perhaps even infinite in length and a long confidence interval is a warning that the instrument provides little information about the treatment. In sum, the confidence intervals under this approach provide clear warnings about the the adherence to IV assumptions (A2) and (A3).

Despite the attractive properties of the exact confidence interval, one drawback is the lack of a closed-form expression and consequently, the computation required to compute the confidence interval. In particular, equation \eqref{eq:CIExact} requires computing $q_{1-\alpha/2}$, the quantile of the null distribution, and doing a very large grid search to find the set of values $\mathcal{F}$ under the null hypothesis $H_0: \tau = \tau_0$. The next section describes a simpler alternative that, unlike the exact method, provides a closed-form expression of the confidence interval.

\subsection{The Almost Exact Method} 
\label{sec:AlmostExact}

The almost exact approach builds on the exact method above by addressing its biggest limitation of computational infeasibility and provides a closed-form expression for the confidence interval. Specifically, we can use ``finite sample asymptotics'' \citep{Hajek:1960,Lehmann:2004} that approximates the exact null distribution in equation \eqref{eq:testStatNull} by considering an asymptotically stable sequence of finite populations $\mathcal{F}$. In the end, we have an asymptotic approximation to the exact confidence interval in equation \eqref{eq:CIExact}, which leads to a closed-form expression for the confidence interval based on a quadratic inequality.  \citet{Hansen:2008b} consider a similar approximation using sample theoretic arguments.

Formally, suppose we have the following estimators for the variances and covariance of $\hat{\tau}_{D}$ and $\hat{\tau}_Y$:
\begin{align*}
\widehat{Var}(\hat{\tau}_{D} | \mathcal{F}, \mathcal{Z}) &=  \frac{1}{n_1(n_1 - 1)} \sum_{i=1}^{n} Z_i \left(D_{i} - \frac{1}{n_1} \sum_{i=1}^{n} Z_iD_{i}\right)^2 \\
&\quad{}+ \frac{1}{n_0(n_0 - 1)} \sum_{i=1}^{n} (1 - Z_i) \left(D_{i} - \frac{1}{n_0} \sum_{i=1}^{n} (1-Z_i)D_{i} \right)^2  \\
\widehat{Var}(\hat{\tau}_{Y} | \mathcal{F}, \mathcal{Z}) &=  \frac{1}{n_1(n_1 - 1)} \sum_{i=1}^{n} Z_i \left(Y_{i} - \frac{1}{n_1} \sum_{i=1}^{n} Z_iY_{i} \right)^2 \\
&\quad{}+ \frac{1}{n_0(n_0 - 1)} \sum_{i=1}^{n} (1 - Z_i) \left(Y_{i} - \frac{1}{n_1} \sum_{i=1}^{n} (1-Z_i)Y_{i} \right)^2 \\
\widehat{Cov}(\hat{\tau}_Y, \hat{\tau}_D | \mathcal{F}, \mathcal{Z}) &= \frac{1}{n_1(n_1 - 1)} \sum_{i=1}^{n} Z_i\left(Y_{i} - \frac{1}{n_1} \sum_{i=1}^{n} Z_iY_{i}\right)\left(D_i - \frac{1}{n_1} \sum_{i=1}^{n} Z_i D_i\right) \\
&\quad{} + \frac{1}{n_0 (n_0 -1)} \sum_{i=1}^{n} (1-Z_i)\left(Y_{i} - \frac{1}{n_0} \sum_{i=1}^{n} (1-Z_i)Y_{i}\right)\left(D_i - \frac{1}{n_0} \sum_{i=1}^{n}(1- Z_i) D_i\right) 
\end{align*}
These are the usual variance estimators for the two-sample mean problems and their properties have been extensively studied under the finite-sample framework; see \citet{Imbens:2015}. In particular, \citet{Imbens:2015} recommends these variance estimators due to their simplicity and attractive properties when extended to infinite-population settings. 

Let  $z_{1-\alpha/2}$ is the $1 -\alpha/2$ denote the quantile for the standard Normal and define $a$, $b$, and $c$as follows.
\begin{align*}
a &= \hat{\tau}_{D}^2 - z_{1-\alpha/2}^2 \widehat{Var}(\hat{\tau}_{D}|\mathcal{F}, \mathcal{Z}) \\
b &= -2 \left(\hat{\tau}_{D} \hat{\tau}_{Y} - z_{1-\alpha/2}^2 \widehat{Cov}(\hat{\tau}_{D}, \hat{\tau}_{Y} | \mathcal{F}, \mathcal{Z}) \right)\\
c &= \hat{\tau}_{Y}^2 - z_{1-\alpha/2}^2 \widehat{Var}(\hat{\tau}_{Y}| \mathcal{F},\mathcal{Z})
\end{align*}
In the Supplementary Materials, we show that the exact confidence interval in equation \eqref{eq:CIExact} is approximately equal to solving the following quadratic inequality based on $a$, $b$, and $c$
\begin{equation} \label{eq:CIExactQuad}
\left\{\tau_0 : P_{H_0}\left(\left| \frac{T(\tau_0)}{S(\tau_0)}\right| \leq q_{1-\alpha/2} | \mathcal{F}, \mathcal{Z}\right)  \right\} \approx  \{\tau_0 : a\tau_0^2 + b\tau_0 + c \leq 0\}
\end{equation}
We stress that this equivalence relation does not rely on assumptions about constant treatment effects nor the exclusion restriction (A3), similar to the exact method. The equivalence relation in \eqref{eq:CIExactQuad}, which we call the almost exact method, allows us to easily compute an approximation to the exact confidence intervals using any standard quadratic inequality solver. In particular, depending on the value of $a$ and the determinant $b^2 - 4ac$, the quadratic inequality can lead to different types of confidence intervals. As an example, if $a > 0$ and $b^2 - 4ac > 0$, which is the only case where the interval is non-empty and finite, a closed-form formula for the confidence interval for $\tau$ using the almost-exact method is
\begin{align}
\label{eq:CIAlmostExactFormula}
\begin{split}
&\frac{\hat{\tau}_D \hat{\tau}_Y - z_{1-\alpha/2}^2  \widehat{Cov}(\hat{\tau}_{D}, \hat{\tau}_{Y} | \mathcal{F}, \mathcal{Z})}{\hat{\tau}_{D}^2 - z_{1-\alpha/2}^2 \widehat{Var}(\hat{\tau}_{D}|\mathcal{F}, \mathcal{Z})} \\
\pm& z_{1-\alpha/2} \frac{ \sqrt{\hat{\Delta} + z_{1-\alpha/2}^2 (\widehat{Cov}^2(\hat{\tau}_{D}, \hat{\tau}_{Y} | \mathcal{F}, \mathcal{Z}) - \widehat{Var}(\hat{\tau}_{D}|\mathcal{F}, \mathcal{Z})\widehat{Var}(\hat{\tau}_{Y}| \mathcal{F},\mathcal{Z})) }}{\hat{\tau}_{D}^2 - z_{1-\alpha/2}^2 \widehat{Var}(\hat{\tau}_{D}|\mathcal{F}, \mathcal{Z})}
\end{split}
\end{align}
where 
\begin{equation} \label{eq:Delta}
\hat{\Delta} =  \hat{\tau}_{Y}^2 \widehat{Var}(\hat{\tau}_{D}|\mathcal{F}, \mathcal{Z}) + \hat{\tau}_{D}^2 \widehat{Var}(\hat{\tau}_{Y}| \mathcal{F},\mathcal{Z}) - \hat{\tau}_D \hat{\tau}_Y \widehat{Cov}(\hat{\tau}_{D}, \hat{\tau}_{Y} | \mathcal{F}, \mathcal{Z})
\end{equation}
The Supplementary Materials details the different solutions to a quadratic equation and corresponding confidence intervals that arise from equation \eqref{eq:CIExactQuad}. 

As we will see in Sections \ref{sec:sim} and \ref{sec:app}, the approximation in equation \eqref{eq:CIExactQuad} works very well, even in situations when the instrument is very weak so that $\tau_D \approx 0$ and $(A2)$ is almost violated. Indeed, the almost exact method, like the exact method, produces infinite confidence intervals and this occurs if $a < 0$, or equivalently,
\begin{equation} \label{eq:Ftest}
\left|\frac{\hat{\tau}_{D}}{\sqrt{\widehat{Var}(\hat{\tau}_{D} | \mathcal{F}, \mathcal{Z})}} \right| \leq z_{1-\alpha/2}
\end{equation}
Equation \eqref{eq:Ftest} is the t-test for testing the strength of the instrument $Z$'s association with $D$ under the null hypothesis $H_0: \tau_D = 0$ and we retain the null of the t-test at $z_{1-\alpha/2}$; see Supplementary Materials for the technical derivations. That is, the almost exact method produces an infinite confidence interval if we cannot reject the null that the instrument is weak, i.e. $H_0: \tau_D = 0$,  at the $\alpha$ level. In short, the almost exact method retains the advantages of the exact method, especially with regards to a weak instrument, but has a closed-form expression that can be easily computed. The only cost to this approach is that it is not exact in finite samples. However, our empirical applications in Section \ref{sec:app1} show that this cost is minimal, especially in samples sizes where the exact method becomes computationally infeasible, while retaining many of the advantages of the exact method.

In sum, both the exact and almost exact approaches to IV inference will produce infinite confidence intervals when an instrument is weak. How should analysts interpret infinite confidence intervals? Is an infinite confidence some type of mistake? An infinite confidence interval is a designed feature of the exact and almost exact approaches. An infinite confidence interval is a warning that the data contain little information \citep[p. 185]{Rosenbaum:2002}. Specifically, it is a warning that the instrument has little effect on the exposure. In fact, the possibility of an infinite confidence interval is a necessary condition for proper inference when an instrument is weak \citep{Dufour:1997}. Investigators may choose to narrow an infinite confidence interval by using prior information. This can be done informally by using IV methods that rely on asymptotic approximations, altering $\alpha$ post-hoc, or using Markov Chain Monte Carlo methods \citep{kleibergen2003bayesian}. We have no objections to this approach, so long as the role of prior information is communicated clearly to the audience. Ideally, the infinite confidence interval would be reported along with any interval using prior information. This will communicate clearly that prior information is needed to produce a non-infinite confidence interval.

\section{Comparison to Traditional Methods of Inference}
\label{sec:CIAsymptotic}
In this section, we discuss more popular modes of inference for IV and provide a comparison to randomization based inferential methods discussed previously. First, the most widely used method of inference depends on an approximation to the Normal distribution and is the basis for inference using TSLS. Specifically, we simply add/subtract the standard error to the point estimate of $\tau$, $\hat{\tau}$, which is also the TSLS estimator, to obtain a $1-\alpha$ confidence interval, 
\begin{equation} \label{eq:CIAsymptotic}
\hat{\tau} \pm z_{1-\alpha/2} \sqrt{ \widehat{Var}(\hat{\tau} | \mathcal{F}, \mathcal{Z})}
\end{equation}
The validity of equation \eqref{eq:CIAsymptotic} relies on $\hat{\tau}$ being approximately normally distributed with mean $\tau$ and standard moment arguments. Specifically, the following must be true: (i) the moment of the product of the structural error term and the instrument is zero, (ii) there is at least one instrument, and (iii) the instrument is sufficiently strong (see Chapter 5 of \citet{Wooldridge:2010} for details). If these conditions hold, the asymptotic approximation in equation \eqref{eq:CIAsymptotic} should be accurate. This approach is the default method of inference for IV in many econometric textbooks \citep{Angrist:2008,Wooldridge:2010} and software (e.g.  the AER R package \citep{kleiber2008AER}). 

Estimating the variance of $\hat{\tau}$ in equation \eqref{eq:CIAsymptotic} can be done using a variety of ways. For example, following \citet{Imbens:2015}, suppose we assume
\[
\begin{pmatrix}
\hat{\tau}_{Y} \\
\hat{\tau}_{D}
\end{pmatrix}
\sim 
N \left( \begin{bmatrix} \tau_{Y} \\ \tau_{D} \end{bmatrix}, \begin{bmatrix} Var(\hat{\tau}_{Y} | \mathcal{F}, \mathcal{Z}) & Cov(\hat{\tau}_{Y}, \hat{\tau}_{D} | \mathcal{F}, \mathcal{Z})
\\ Cov(\hat{\tau}_{Y}, \hat{\tau}_{D} | \mathcal{F}, \mathcal{Z}) & Var(\hat{\tau}_{D} | \mathcal{F}, \mathcal{Z}) \end{bmatrix} \right)
\]
Then, the Delta method can be used to derive an approximation of the variance of $\hat{\tau}$
\begin{equation} 
Var(\hat{\tau} | \mathcal{F}, \mathcal{Z}) \approx \frac{Var(\hat{\tau}_{Y} | \mathcal{F}, \mathcal{Z})}{\tau_{D}^2} + \frac{\tau_{Y}^2 Var(\hat{\tau}_{D} | \mathcal{F}, \mathcal{Z})}{\tau_{D}^4} - \frac{2 \tau_{Y} Cov(\hat{\tau}_{Y},\hat{\tau}_{D} | \mathcal{F}_{n}, \mathcal{Z})}{ \tau_{D}^3} \label{eq:varEcon}
\end{equation}
and plugging an estimate of this variance into equation \eqref{eq:CIAsymptotic} results in the following $1-\alpha$ confidence interval for $\tau$
\begin{equation} \label{eq:CIImbens}
\hat{\tau} \pm z_{1-\alpha/2} \sqrt{ \frac{\widehat{Var}(\hat{\tau}_{Y} | \mathcal{F}, \mathcal{Z})}{\hat{\tau}_{D}^2} + \frac{\hat{\tau}_{Y}^2 \widehat{Var}(\hat{\tau}_{D} | \mathcal{F}, \mathcal{Z})}{\hat{\tau}_{D}^4} - \frac{2 \hat{\tau}_{Y} \widehat{Cov}(\hat{\tau}_{Y},\hat{\tau}_{D} | \mathcal{F}_{n}, \mathcal{Z})}{ \hat{\tau}_{D}^3} }
\end{equation}
With some algebra, one can show that the confidence interval in equation \eqref{eq:CIImbens} is equivalent to the confidence interval used for the TSLS estimator; see Supplementary Materials for details. For simplicity, we'll refer to this approach as the TSLS or the Delta method.

Another approach to estimating the variance in equation \eqref{eq:CIAsymptotic} is by treating $\hat{\tau}_{D}$ as fixed so that the only random component of $\hat{\tau}$ is $\hat{\tau}_{Y}$. That is, assume that the effect of $Z_i$ on $D_i$ is known without error. This approach leads us to a variance of $\hat{\tau}$ which is the variance of $\hat{\tau}_{Y}$ divided by $\hat{\tau}_D$ and the resulting $1- \alpha$ confidence interval formula in \eqref{eq:CIAsymptotic} is
\begin{equation} \label{eq:CIBloom}
\hat{\tau} \pm z_{1-\alpha/2} \sqrt{\frac{\widehat{Var}(\hat{\tau}_{Y} | \mathcal{F}, \mathcal{Z})}{\hat{\tau}_{D}^2}}
\end{equation}
This simple approximation was proposed by \citet{Bloom:1984} who suggested it in the context of program evaluation under noncompliance, and it is widely used in the program evaluation literature judging from recent citation patterns. Econometrics, statistics, and Mendelian randomization also utilize this approximation \citep{Heckman:1998c, Yang:2013, bowden_assessing_2016}; in particular, Mendelian randomization studies with summary data refers to this approximation as No Measurement Error (NOME) assumption \citep{bowden_assessing_2016}. Hereafter, we refer to this method of inference as the Bloom method.

The two asymptotic variance estimates based on the Delta method and the Bloom method are related as follows. In simple terms, the variance from the Bloom method is exactly the first term of the variance from the Delta method in equation \eqref{eq:varEcon}. More specifically, if we denote $\widehat{Var}(\hat{\tau} | \mathcal{F}, \mathcal{Z})_{Bloom}$ to be the variance estimate used in \eqref{eq:CIBloom}, i.e. $\widehat{Var}(\hat{\tau} | \mathcal{F}, \mathcal{Z})_{Bloom} = \widehat{Var}(\hat{\tau}_{Y} | \mathcal{F}, \mathcal{Z}) / \hat{\tau}_{D}^2$ and $\widehat{Var}(\hat{\tau} | \mathcal{F}, \mathcal{Z})_{Delta}$ to be the the variance estimate used in \eqref{eq:CIImbens}, the two variance estimates are related by a factor $C > 0$
\[
\widehat{Var}(\hat{\tau} | \mathcal{F}, \mathcal{Z})_{Delta}  = \widehat{Var}(\hat{\tau} | \mathcal{F}, \mathcal{Z})_{Bloom} C,
\]
and $C$ is defined as
\begin{equation}  \label{eq:CVar}
C = \left( 1 +\frac{\hat{\tau}_{Y}^2 \widehat{Var}(\hat{\tau}_{D} | \mathcal{F}, \mathcal{Z})}{\hat{\tau}_{D}^2\widehat{Var}(\hat{\tau}_{Y} | \mathcal{F}, \mathcal{Z})} - \frac{2 \hat{\tau}_{Y}\widehat{Cov}(\hat{\tau}_{Y},\hat{\tau}_{D} | \mathcal{F}, \mathcal{Z})}{\hat{\tau}_{D}\widehat{Var}(\hat{\tau}_{Y} | \mathcal{F}, \mathcal{Z})}  \right)
\end{equation} 
When $C > 1$, $\widehat{Var}(\hat{\tau} | \mathcal{F}, \mathcal{Z})_{Delta}$ is larger than $ \widehat{Var}(\hat{\tau} | \mathcal{F}, \mathcal{Z})_{Bloom} $ and the Delta confidence interval is larger than the Bloom confidence interval. In contrast, $C < 1$ would imply the opposite and the Delta confidence interval would be smaller than the Bloom confidence interval. We can show that $C < 1$ occurs if and only if
\[
|\hat{\tau}| < \left| \frac{2\widehat{Cov}(\hat{\tau}_{Y},\hat{\tau}_{D} | \mathcal{F},\mathcal{Z})}{\widehat{Var}(\hat{\tau}_{D} | \mathcal{F}, \mathcal{Z})} \right|
\]
Also, given any $\hat{\tau}_Y$, $\widehat{Var}(\hat{\tau}_{Y} | \mathcal{F}, \mathcal{Z})$, and $\widehat{Cov}(\hat{\tau}_{Y},\hat{\tau}_{D} | \mathcal{F}, \mathcal{Z})$, as $\left| \hat{\tau}_{D}/\sqrt{\widehat{Var}(\hat{\tau}_{D} | \mathcal{F}, \mathcal{Z})} \right|$ increases, i.e. as the instrument becomes stronger based on the t-test measure of instrument strength in equation \eqref{eq:Ftest}, or if $\hat{\tau}_{D}$ has little variability so that $\sqrt{\widehat{Var}(\hat{\tau}_{D} | \mathcal{F}, \mathcal{Z})}$ decreases to zero and the t-test increases, $C$ will get closer to $1$. This is because the denominator in equation \eqref{eq:CVar} gets larger, effectively making $C \approx 1$. Ultimately, this suggests that for strong instruments, the difference between the two variance estimates, and consequently their respective confidence intervals in equations \eqref{eq:CIBloom} and \eqref{eq:CIImbens}, will be negligible.

However, we emphasize an important caveat for both approaches: both depend on the assumption of asymptotic Normality of $\hat{\tau}$. In fact, if the instrument is weak so that $\hat{\tau}_D \approx 0$, $\hat{\tau}$ will be far from Normal and the asymptotic confidence interval via equation \eqref{eq:CIAsymptotic} will be highly misleading: see \citet{Staiger:1997} and \citet{Stock:2002} for details. In contrast, the exact and the almost exact confidence intervals in equations \eqref{eq:CIExact} and \eqref{eq:CIExactQuad}, respectively, can  provide honest coverage for $\tau$ when the instrument is weak.  These methods do not rely on the Normality assumption and instead, use the distribution of $\mathbf{Z}$ from the experimental design as the starting point for inference on $\tau$. Furthermore, the derivation of inference for methods based on equation \eqref{eq:CIAsymptotic} typically requires additional identifying assumption (A3), which is akin to the fact the moment of the product of the structural errors and the instrument is zero. 

We can also compare the Delta method confidence interval (i.e. the TSLS confidence interval) and the almost exact interval as follows. One can rewrite the Delta method confidence interval in equation \eqref{eq:CIImbens} as 
\begin{align} 
\label{eq:CIImbensVSCIExactQuad}
\hat{\tau} \pm z_{1-\alpha/2} \frac{\sqrt{\hat{\Delta}}}{\hat{\tau}_{D}^2} 
\end{align}
where $\hat{\Delta}$ was defined in equation \eqref{eq:Delta} of the almost exact interval. Thus, if the almost exact interval produces finite intervals, i.e. if the instrument is sufficiently strong at the $\alpha$ level (see Section \ref{sec:AlmostExact} for details), the Delta method confidence intervals and the almost exact interval looks similar, but with notable differences in the center and scaling between the two intervals.

Further comparisons of the expressions for different confidence intervals under the same notation provides some insight into the relationship between the traditional confidence intervals based on a point estimator plus and minus the variance in equation \eqref{eq:CIAsymptotic} versus confidence intervals based on randomization inference. The Bloom confidence interval in equation \eqref{eq:CIBloom} is the crudest, yet simplest approximation of inference for $\tau$. The Delta confidence interval in equation \eqref{eq:CIImbens} offers a better approximation than the Bloom confidence interval by incorporating the variability of $\hat{\tau}_D$ and this is reflected by additional scaling terms to the right of $Var(\hat{\tau}_{Y} | \mathcal{F}, \mathcal{Z}) / \hat{\tau}_D^2$ in equation \eqref{eq:varEcon}. The almost exact interval in equation \eqref{eq:CIExactQuad} improves upon the Delta confidence interval by considering the case when $\hat{\tau}_{D} \approx 0$. Consequently, as seen in equation \eqref{eq:CIImbensVSCIExactQuad}, we have slight differences in centering and scaling between the almost exact method and the Delta confidence interval. Finally, the exact interval in equation \eqref{eq:CIExact} provides the exact confidence interval for $\tau$, but without a closed-form solution.


Next, we highlight some relationships between the almost exact interval, the Anderson-Rubin confidence interval \citep{anderson_estimation_1949}, and the interval suggested by Fieller \citep{fieller_some_1954,burgess_review_2015}. First, the almost exact interval is similar to the Anderson-Rubin confidence interval which is popular in the weak instrument literature in econometrics, with asymptotic equivalence when (i) $n$ is large and (ii) homoskedastic variance is assumed; see the Supplementary Materials for details. However, the motivation for the Anderson-Rubin confidence interval, like the TSLS-type confidence intervals, typically relies on structural modeling assumptions or moment conditions; see \citet{Stock:2002} for one examples. Instead, we motivated the almost exact method using the randomization-inference framework where we only made assumptions on the randomization distribution of the instrument $Z_i$ that was part of the design of the experiment. Also, the asymptotic equivalence between the Anderson-Rubin confidence interval and the almost exact interval suggests that the Anderson-Rubin confidence interval is very similar to the exact interval in \eqref{eq:CIExact}. Second, the interval by Fieller is equivalent to the almost exact interval; specifically, equation (5) of \citet{fieller_some_1954} is identical to our quadratic equation of the almost exact interval in equation \eqref{eq:CIExactQuad}. However, similar to the Anderson-Rubin confidence interval whose derivation typically required moment assumptions, Fieller derived his interval under stronger assumptions of (i) bivariate Normality of $\hat{\tau}_Y$ and $\hat{\tau}_{D}$ and (ii) independent estimates of the covariances of the bivariate Normal and the mean of the bivariate Normal. In contrast, our almost exact interval is derived based on the distribution of the instrument $Z_{i}$, which is inherent by the experimental design. Indeed, the fact that our randomization-based intervals achieve these equivalence relationships with weaker assumptions demonstrates the strength of the randomization-inference framework to conduct inference in IV settings.

\section{Extensions}

\subsection{Alternative Test Statistics}

As we noted above, we have thus far considered exact approaches to inference that do not assume constant treatment effects at the unit level through the use of averages. However, results based on averages may be misleading when the tails of the outcome distribution are not well-behaved. One additional advantage of the exact approach is that one can easily use other statistics that may be more robust.  For example, one can use rank-based test statistics, which are robust in the presence of outliers and heavy-tailed distributions. Next, we briefly demonstrate how under the exact approach, investigators may avoid the use of averages. Here, we must adopt a model of effects or a model for how units respond to treatment. \citet{Rosenbaum:1999a} outlines a model of effects where the effect of encouragement on response is proportional to its effect on the treatment dose received:
\begin{equation}
\label{eq:mod.eff}
Y_i^{\left(1,D_i^{(1)}\right)} -Y_{i}^{\left(0,D_i^{(0)}\right)} = \beta(D_i^{(1)} - D_{i}^{(0)}).
\end{equation}
\noindent If this model is holds then observed responses are related to observed doses through the following equation
\begin{equation*}
Y_{i} - \beta D_{i} = Y_i^{\left(1,D_i^{(1)}\right)} - \beta D_i^{(1)} = Y_{i}^{\left(0,D_i^{(0)}\right)} - \beta D_{i}^{(0)}.
\end{equation*}
Under this model of effects, the response will take the same value regardless of the value of $Z_{i}$, which makes this model of effects consistent with the exclusion restriction. Informally, the exclusion restriction implies that instrument assignment $Z_{i}$ is related to the observed response $Y_{i} = Z_{i} Y_i^{\left(1,D_i^{(1)}\right)} + (1 - Z_{i}) Y_{i}^{\left(0,D_i^{(0)}\right)}$ only through the realized dose of the treatment $D_{i}$. That is true here since $Y_{i} - \beta D_{i}$ is a constant that does not vary with $Z_{i}$. Under this model of effects, the treatment effect varies from unit to unit based on the level of $D_i$ as measured by $(D_i^{(1)} - D_{i}^{(0)})$.  If the unit received no dose, then $Y_i^{\left(1,D_i^{(1)}\right)} -Y_{i}^{\left(0,D_i^{(0)}\right)} = \beta(D_i^{(1)} - D_{i}^{(0)})=0$. 

Exact inference about $\beta$ using rank based methods involves no new principles except that we must invoke a model of effects.  Note that we do not assume the effect is constant, here, the treatment effect varies from unit to unit based on the level of $D_i$. However, we must invoke a model for treatment response.  Here, we assume that the effect of encouragement on response is proportional to its effect on the treatment dose received. Thus, alternative tests statistics may require an additional assumption about response to treatment; see \citet{Keele:2016a} for an exception when the outcome is binary. To test the sharp null hypothesis, $H_0: \beta = \beta_0$, using rank based methods, we use the observed quantity $Y_{i} - \beta_0 D_{i} = W_i$ as a set of adjusted responses. For example, to use the Wilcoxon rank sum statistic, we rank $|W_i|$ from 1 to $N$, and take the sum of the ranks for which $Z_i = 1$ to form the test statistic $W_{\beta_0}$. Comparing $W_{\beta_0}$ to the randomization distribution for Wilcoxon rank sum statistic provides an exact $p$-value for a test of the sharp null hypothesis that $H_0: \beta = \beta_0$. A 95\% confidence interval for the treatment effect is formed by inverting the test, or testing a series of hypotheses $H_0: \beta = \beta_0$ and retaining the set of values of $\beta_0$ not rejected at the 5\% level \citep{Rosenbaum:2002}.  Rank-based methods are just one alternative. Using the adjusted responses, one could use a variety of test statistics.  For example, one could test for differences in higher-order moments of the adjusted responses.

\subsection{Sensitivity Analysis} \label{sec:sens}

When IV is used outside of randomized experiments with noncompliance, assumption (A1) no longer holds by design.  Typically, analysts assume that assignment to encouragement (the instrument) is as-if random or as-if random conditional on observed covariates. In applications of this type, exact and almost exact methods allow a sensitivity analysis that assesses how departures from random assignment of the instrument might alter our conclusions. \citet{Rosenbaum:2002} demonstrates that with exact methods, the analyst can place sharp bounds on the $p$-value from the test of the sharp null given the effect of a hypothetical confounder on the probability of being assigned to encouragement. 

Rosenbaum uses $\Gamma$ to represent the odds of being assigned to treat $(Z_i=1)$ for a unit as a function of a possible binary unobserved confounder. When $\Gamma = 1$, then the units do not differ in their odds of treat assignment as a function of the unobserved confounder.  This is true by design when $Z_i$ is randomly allocated, but may not be true in a natural experiment or observational study. To conduct a sensitivity analysis, the analyst uses values of $\Gamma$ larger than one to place bounds on the $p$-value from the test of the sharp null. For example, if we find that the bounds on the $p$-value exceed 0.05 when $\Gamma = 1.05$ this suggests that a very slight departure from randomization of the instrument might overturn the conclusion from the study.  If, on the other hand, we find that the $p$-value from our study exceeds 0.05 when $\Gamma$ is greater than say 4, this suggests that unless the departure from randomization of the instrument is fairly substantial our conclusions would still hold.  These methods can be easily extended to the almost exact approach. See \citet{Baiocchi:2010}, \citet{Keele:2014b}, and \citet{kang2016full} for examples of this form of sensitivity analysis applied to studies with instrumental variables.

\subsection{Covariate Adjustment}

Next, we consider covariate adjustment. Analysts often include baseline covariates in IV analyses.  In randomized trials, covariates may increase precision, and in observational studies adjusting for covariates may be used to remove biases due to nonrandom assignment of instrument status. Under the asymptotic approaches in Section \ref{sec:CIAsymptotic}, such as TSLS, analysts can simply include baseline covariates in the linear regression models for the outcome and the treatment. The exact and almost exact approaches, however, do not preclude the use of baseline covariates. \citet{Rosenbaum:2002} outlines a general method for covariate adjustment with exact methods. Under his approach, exact methods are applied to the residuals from a model where the outcome has been regressed on baseline covariates.  This method directly extends to the almost exact approach. Alternatively, as noted above covariate adjustment may also be applied via matching and then either exact and almost exact methods are used. \citet{Baiocchi:2010}, \citet{Keele:2014b}, and \citet{kang2016full} detail the use of both exact and almost exact methods in IV analyses where matching is used for covariate adjustment.

\subsection{Multi-valued instruments and treatments}

In many IV applications, either the instrument $Z_i$ or actual treatment exposure $D_i$ may not be binary. For example, $D_i$ may record self-selected treatment dosage or $Z_i$ may be some continuous level of encouragement to take the treatment.  When $D_i$ is multi-valued, the randomization inference approach laid out in this paper remains the same and still provides correct confidence intervals without any modification. Note, however, that a multi-valued treatment does change the interpretation of the estimand $\tau$; see \citet{angrist_two-stage_1995} and \citet{kang2016full} for examples. When instruments are multi-valued, randomization inference can remain valid by using two strategies. First, investigators can simply create a binary IV from the continuous version. \citet{Angrist:1994} show that an IV estimate with a binary instrument and a continuous instrument converge to weighted averages of treatment effects, where the more `compliant' subjects obtain greater weight. Second, \citet{Baiocchi:2012} demonstrates how dichotomizing multi-valued instruments may be avoided through the use of non-bipartite matching.

\section{Simulation}  
\label{sec:sim}

We now present a simulation study to compare the properties of the different inferential methods. The study evaluates the confidence interval coverage of three methods of IV inference: the almost exact method, the TSLS method, and the Bloom method. We do not include the exact method in the simulation since it is guaranteed to always have nominal coverage in finite samples while the other three methods are approximations of the finite sample behavior. The simulation considers one-sided compliance under a finite sample and we evaluate the coverage rate for each method as the proportion of the compliers, $\tau_D$, varies. First, we sample $Z_i$ from a Bernoulli (0.5) distribution. To simulate the proportion of compliers, $\tau_D$, we set the compliance rate to $\pi$ and sample units from a uniform distribution. Let $P_i$ denote the compliance class, which only include compliers and never takers in the one-sided compliance setting. We designate a unit as complier, $co$, if the draw from the uniform distribution is less than $ \pi$. The outcomes $Y_i$ follow a Normal distribution with mean $\kappa + \gamma I(P_i = co)$ and variance $\sigma^2$ where $co$ indicates that a unit is a complier and $I(\cdot)$ and is the indicator function. Under this model, $\gamma$ is the effect on $Y_i$ for $D_i = 1$ versus $D_i = 0$ for compliers. In the simulation, we set $\kappa = \gamma = \sigma^2 = 1$. This type of simulation setup is not new and follows closely the simulation design in \citet{guo2014using}.

In the simulations, we varied the the compliance rate using an interval of 5\%, 10\%, 25\%, 50\%, 75\%, and 90\%.  This implies that assignment to $Z_i=1$ results in 5\% to 90\% of units being exposed to $D_i=1$. Also, to study the behavior at low compliance rates, we also add one additional compliance rate based on our discussion in Section \ref{sec:AlmostExact}. Specifically, based on equation \eqref{eq:Ftest}, the almost exact method of inference will return an infinite confidence interval if 
\begin{equation}
\hat{\tau}_{D} \leq \frac{z_{1-\alpha/2}^2}{n + z_{1-\alpha/2}^2} 
\label{eq:ci.in}
\end{equation} 
where we set $\widehat{Var}(\hat{\tau}_{D} | \mathcal{F}, \mathcal{Z}) = \hat{\tau}_{D} (1 -\hat{\tau}_D)/n$ for one-sided compliance. In the simulations, we set the sample size to 100 and $\alpha = 0.05$, so under equation \eqref{eq:ci.in}, infinite confidence intervals will occur when the compliance rate is approximately 1.9\%. Therefore, for one set of simulations, we set the compliance rate to 1.9\%. We expect the coverage rates for confidence intervals based on Normal approximations will tend to have incorrect coverage when the compliance rate falls around 1.9\%. We repeat our simulation 5000 for each method in each scenario. For each simulation, we record the 95\% coverage rate for each method.

The results from the simulations are in Tables~\ref{tab:sim} and~\ref{tab:sim_length}. First, we observe that for the almost exact method, the coverages are at the nominal rate for any level of compliance. That is, even when less than two percent of the units are compliers, the almost exact methods maintains 95\% coverage. As such, the almost exact approximation appears to be quite accurate, even when the sample size is 100. Later, in an empirical example, we compare the almost exact method to the exact method.

For the asymptotic methods, both perform very poorly at the lowest levels of compliance. When the compliance rate is less than two percent, the coverage rates for both methods fail to reach 50\%. When the compliance rate is 5\%, the Bloom method fails to have a 90\% coverage rate, while the TSLS method is close to the nominal coverage rate at 94\%.  However, for compliance rates above 75\% both the Bloom and the TSLS method have the correct coverage.  Moreover, the Bloom method appears to be an accurate approximation to the TSLS method.  Once the compliance rate is 25\% or higher the two methods have nearly identical coverage rates. 

\begin{table}[ht]
\centering
\caption{Coverage rates of confidence intervals from almost exact, Bloom, and the TSLS method.}
\label{tab:sim}
\begin{tabular}{lccccccc}
\toprule
 Compliance Rate & 1.9\% & 5\% & 10\% & 25\% & 50\% & 75\% & 90\% \\ 
\midrule
Almost Exact & 0.950 & 0.944 & 0.945 & 0.947 & 0.955 & 0.941 & 0.948 \\ 
Bloom & 0.477 & 0.885 & 0.947 & 0.956 & 0.967 & 0.953 & 0.956 \\ 
TSLS  & 0.503 & 0.934 & 0.996 & 0.978 & 0.964 & 0.945 & 0.949 \\   
\bottomrule
\end{tabular}
\end{table}

Table~\ref{tab:sim_length} looks at the median length of the confidence intervals. At low compliance rates, we observe the confidence interval length for the almost exact method is either infinite or very large, reflecting the uncertainty that is inherent with low compliance and theoretically achieving the infinite length requirement laid out in \citet{Dufour:1997} for a weak instrument. In contrast, the Bloom and the TSLS methods tend to have large intervals as the compliance rate decreases, but fails to achieve coverage. In fact, by design, the Bloom and the TSLS methods can never have infinite confidence intervals while the almost exact can create infinite confidence intervals. Specifically, in our simulations, we noticed that 97.6\% of the 5000 simulated confidence intervals from the almost exact method were infinite when the compliance rate was 1.9\%, 93.5\% when the compliance rate was 5\%, 26.2\% when the compliance rate was 10\%, and 0.1\% when the compliance rate was 25\%. However, the length of the almost exact method confidence intervals conveys important information even when they are not infinite. For example, when compliance is 10\% the asymptotic confidence intervals are very similar, while the almost exact intervals are nearly 3 times longer. Thus these intervals better convey the true level of statistical uncertainty. Finally, we also recorded the average point estimate for the almost exact method. For any of the compliance rates of 5\% or higher, the bias associated with the almost exact method was 1\% or less. When the compliance rate was 1.9\%, the almost exact point estimate was too small by 32\%.

\begin{table}[ht]
\centering
\caption{Median length of confidence intervals from almost exact, Bloom, and the TSLS method.}
\label{tab:sim_length}
\begin{tabular}{lccccccc}
\hline
 Compliance Rate & 1.9\% & 5\% & 10\% & 25\% & 50\% & 75\% & 90\% \\ 
\hline
Almost Exact & $\infty$        & $\infty$        & 28.962 & 4.934 & 1.766 & 1.067 & 0.942 \\ 
Bloom           & 28.510  & 18.561 &   8.956 & 4.060 & 1.768 & 1.097 & 0.965 \\ 
TSLS             & 30.697  & 20.695 &   9.493 & 4.044 & 1.683 & 1.053 & 0.935 \\   
\hline
\end{tabular}
\end{table}

Next we include one additional simulation to convey how the almost exact method approximates the exact interval. Recall that the almost exact method depends on the three terms $a$, $b$, and $c$ defined in equation \eqref{eq:CIExactQuad}. These terms effectively act as correction factors that produce a confidence interval that approximates the exact interval. Specifically, the almost exact confidence interval depends on $a$ and $b^2 - 4ac > 0$: as the value of these two terms increase the width of the almost exact interval decreases, and it will be finite when $a>0$ and $b^2 - 4ac > 0$. Note that  $c$ plays a relatively small role in the width of the interval, since it only depends on $\hat{\tau}_{Y}$ while both $a$ and $b$ depend on $\hat{\tau}_{D}$.

\begin{figure}
\centering
\includegraphics[scale=.65]{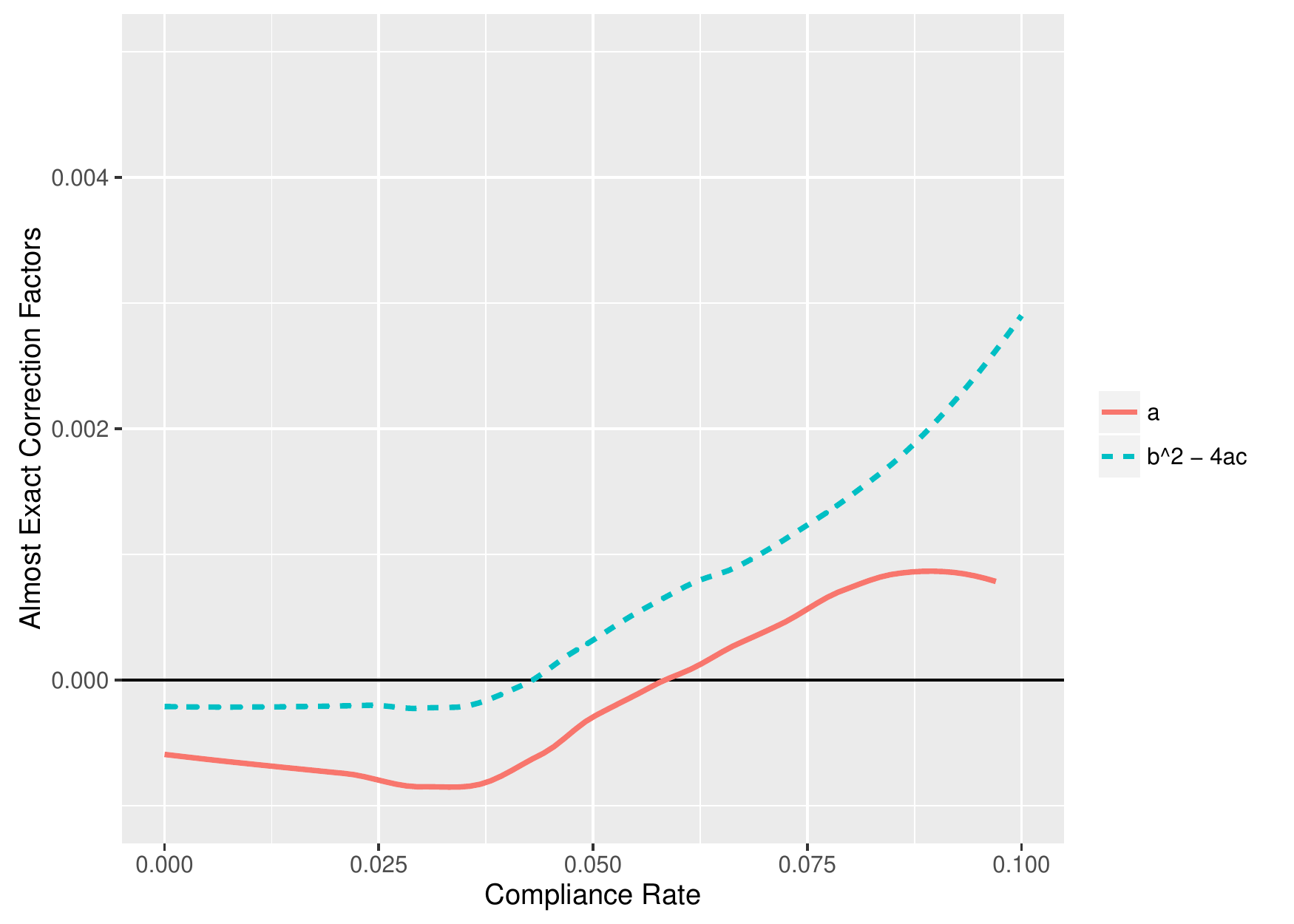}
\caption{Almost exact correction factors $a$ and $b^2 - 4ac$ plotted against compliance rates in simulated data. Almost exact interval will be finite when both values are greater than zero. In general, the width of the almost exact interval shrinks as these factors increase. Note: $a = \hat{\tau}_{D}^2 - z_{1-\alpha/2}^2 \widehat{Var}(\hat{\tau}_{D}|\mathcal{F}, \mathcal{Z})$, $b = -2 \left(\hat{\tau}_{D} \hat{\tau}_{Y} - z_{1-\alpha/2}^2 \widehat{Cov}(\hat{\tau}_{D}, \hat{\tau}_{Y} | \mathcal{F}, \mathcal{Z}) \right)$, and $c=\hat{\tau}_{Y}^2 - z_{1-\alpha/2}^2 \widehat{Var}(\hat{\tau}_{Y}| \mathcal{F},\mathcal{Z})$}
\label{fig:sim_plot}
\end{figure}

In the second simulation, we used the same data generating process and simulation parameters as the first simulation, except we varied the compliance rate from 1\% to 10\% in increments of .001. For each compliance rate, we ran 1000 simulations and recorded the average values of $a$, $b$, and $c$. In Figure~\ref{fig:sim_plot}, we plot smoothed values of $a$ and $b^2 - 4ac$ against the compliance rate. When the compliance rate is less than 2.5\%, both terms are essentially flat. As the compliance rate increases, the correction terms increase which narrows the almost exact interval.  We observe that while the term $b^2 - 4ac$ exceeds zero once compliance rates are around 4\%, both terms do not exceed zero until the compliance rate is approximately 6\%. Next, we use three different empirical applications to highlight different aspects of these inferential methods.

\section{Applications} 
\label{sec:app}

\subsection{Application 1: The Green Jobs and Health Care (GJ-HC) Intervention}
\label{sec:app1}

As part of a comprehensive economic stimulus package funded under the 2009 American Recovery and Reinvestment Act (ARRA), the U.S. Department of Labor awarded a series of grants to promote training for employment in energy efficiency, renewable energy, and health care. Grants were awarded to four sites across the United States. At two sites, additional training was offered on topics such as the installation of solar and wind power systems. At the other two sites, additional training was offered in the health care sector. These new training initiatives were subject to evaluation in the Green Jobs and Health Care (GJ-HC) Impact Evaluation \citep{Copson:2015,Martinson:2015}. 

At each site, participants were randomized to either participation in the new training programs, i.e. treatment, or to the existing training programs available---control. At all four of the sites, some trainees who were assigned to the new training initiatives selected not to participate. However, in the study design, those randomized to the standard training condition could not access the treatment.  Thus noncompliance was one-sided in this application. The primary outcomes were earnings and employment status.  Here, we focus on the employment status outcome which was measured through a survey after trainees completed their course of study. We use a binary outcome measure which asked if the participants had been employed at anytime since ending the training program. 

We conducted a separate analysis for each site given that the content of the training programs varied significantly across the four sites. Table~\ref{tab:app.out} contains the total number of participants in the randomized intervention at each site along with the compliance rate. At all four sites, compliance with assigned treatment status was high. The lowest level of compliance was at Site 1, where 62.1\% of those randomized to treatment participated.  At the other three sites, participation among those assigned to treatment exceeded 75\%.  

In the GJ-HC application, the sample sizes are small enough that exact methods are feasible.  As such, we compare results from an exact method developed in \citet{Keele:2016a} to the almost exact method. Table~\ref{tab:app.out} contains point estimates and 95\% confidence intervals using both exact methods and the approximation to the exact method outlined in Section~\ref{sec:AlmostExact}. First, the approximation clearly improves as the sample size increases. Site 4 has the largest sample size with 719 participants, and the exact and almost exact methods provide essentially identical results. For Site 4, the exact 95\% confidence interval is $-0.05$--$0.06$, and the almost exact 95\% confidence interval is $-0.05$--$0.05$.  However, the computation time for the exact method was over 8 minutes on a desktop with a 4.0 GHz processor and 32.0 GB RAM.  The almost exact routine is essentially instantaneous, since it is based on a closed-form solution. As such, the almost exact method provides an accurate approximation to the exact results, but requires very little computing power. Site 2 has the smallest sample size, so we might expect the discrepancy between the exact and almost exact confidence intervals to be largest for the analysis of this training site. Here, the exact 95\% confidence interval is $[-0.06, 0.29]$ and the almost exact 95\% confidence interval is $[-0.07, 0.19]$. The exact confidence interval, then, is longer as it exactly reflects finite sample uncertainty, and in this case exact methods require very little computation time. These results suggest that analysts should use exact methods when sample sizes are smaller.

\begin{table}[htbp]
\centering
\caption{Point estimates and confidence intervals for exact and almost exact methods in the GJ-HC data.}
\label{tab:app.out}
\begin{tabular}{lcccc}
\hline
 & Site 1 & Site 2 & Site 3 & Size 4 \\
\hline
Hodges-Lehmman Point Est. &0.050 & 0.060 & 0.084 & -0.003 \\ 
Almost Exact 95\% CI  &[-0.06, 0.16] & [-0.07, 0.19] & [0.02, 0.15] & [-0.05, 0.05] \\ 
Exact 95\% CI  &[-0.05, 0.19] & [-0.06, 0.29] & [0.02, 0.17] & [-0.05, 0.06] \\ 
Computation Time in Minutes  &0.04 & 0.01 & 0.63 & 8.9 \\ 
N  & 318 & 169 & 546 & 719 \\ 
Compliance Rate  & 62.1\% & 79.3\% & 79.9\% & 83.9\% \\ 
\hline
\end{tabular}
\end{table}

\subsection{Application 2: A Get-Out-The-Vote Intervention}
\label{sec:app2}

One literature in political science studies methods for increasing voter turnout through the use of randomized field experiments. This research both focuses on the effectiveness of various get-out-the-vote methods and tests social psychological theories about voters \citep{Green:2013}. One entry in this literature focused on the effectiveness of door-to-door canvassing where volunteers knock on doors urging people to vote in an upcoming election \citep{Green:2003a}.  In this study, the researchers conducted six separate field experiments in the following cities: Bridgeport, Columbus, Detroit, Minneapolis, Raleigh, and St. Paul in November 2001.  In each city, households were randomized to either receive face-to-face contact from local staffers encouraging them to vote, i.e. treatment, or were not contacted, i.e. control. Many of the households randomized to the treatment were not available for the face-to-face message encouraging them to vote.  While the intention-to-treat (ITT) effects are easily estimable, in this context, one might argue that IV estimates are of greater interest, since these reveal the causal effect of actually receiving the get-out-the-vote message. In the original analysis, the analysts estimated complier effects using asymptotic approximations for the variance estimates \citep{Green:2003a}.

The sample sizes for these experiments, however, make using exact methods computationally intensive. For example, the experiment in St. Paul had 2146 participants. When we attempted to obtain exact results on a desktop with a 4.0 GHz processor and 32.0 GB RAM, computation stopped after 345 minutes due to the fact that the computer had run out of memory.  Thus, even in fairly modest sample sizes, exact methods may be infeasible.  Here, we compare the almost exact results to results based on the Delta/TSLS method and the Bloom approximation. Table~\ref{tab:local} contains the point estimates and confidence intervals for these three different methods. All three methods produce essentially identical results.  Despite the fact that the compliance rates are, at times, less than 15\%, the larger sample sizes ensure that all three methods produce identical inferences.

\begin{sidewaystable}
\centering
\caption{Point estimates and confidence intervals for exact and almost exact methods in the voting intervention data.}
\label{tab:local}
\begin{tabular}{lcccccc}
\hline
 & Bridgeport & Columbus & Detroit & Minneapolis & Raleigh & St. Paul \\ 
\hline

Point Estimate & 0.163 & 0.105 & 0.083 & 0.104 & -0.020 & 0.138 \\ 
Almost Exact 95\% CI & [0.052, 0.274]  & [-0.061, 0.270]& [-0.007, 0.174]& [-0.071, 0.280]& [-0.081, 0.039]& [0.013, 0.263] \\ 
TSLS  95\% CI & [0.053, 0.273] & [-0.059, 0.269] & [-0.007, 0.173]& [-0.070, 0.279]& [-0.080, 0.039]& [0.014, 0.262] \\  
Bloom 95\% CI & [0.051, 0.275]& [-0.060, 0.269] & [-0.007, 0.173]& [-0.070, 0.279] & [-0.080, 0.039]& [0.013, 0.263] \\
N & 1650 & 2424 & 4954 & 2827 & 4660 & 2146 \\ 
Compliance Rate & 28.9\% & 14.0\% & 30.7\% & 18.5\% & 45.2\% & 33.1\% \\ 

\hline
\end{tabular}
\end{sidewaystable}

\subsection{Application 3: Rainfall as an Instrument for Voter Turnout}


As we noted earlier, instruments have a long history of use outside of randomized trials. Here, instruments are used as a type of natural experiment, where the  instrument is a haphazard nudge or ``encouragement'' to treatment exposure. The likelihood of weak instruments tends to be higher when instruments are used outside of randomized evaluations. As such, the utility of exact or almost exact methods is likely greater when instruments are used in observational studies. Here, we conduct a re-analysis of \citet{Hansford:2010} to explore whether almost exact methods may be useful in an observational study with an instrument. 

\citet{Hansford:2010} use deviations from average rainfall on election day as an instrument for voter turnout to estimate the causal effect of voter turnout on vote share in U.S. elections. Using this instrument, they find that higher turnout tends to help Democratic candidates. The original analysis spanned all presidential elections in non-Southern counties from 1948 to 2000. Here, we investigate the possibility that the strength of rainfall as an IV for voter turnout perhaps declined over time.  That is, changes in transportation patterns over time might weaken the effect of rainfall on turnout, since voters may be less affected by rain when they are able to drive to the polls. In our analysis, we conduct three separate analyses.  The first analysis uses all presidential elections from 1976 to 2000.  The second uses all presidential elections from 1980 to 2000, and the third uses all presidential elections from 1984 to 2000. For every analysis, we have close to 10,000 observations or more.  Thus, uncertainty in the IV estimate will be largely driven by the strength of the instrument, instead of the sample size. We used the almost exact method, the Bloom method and TSLS, which was the method used in the original analysis, for interval estimation. 

Table~\ref{tab:rain} contains the results from the analysis. When we analyze the elections from 1976 to 2000, all three methods return 95\% confidence intervals that are quite similar.  The almost exact method does have wider confidence intervals, but the difference is relatively small. The almost exact 95\% confidence interval is $[-0.91, -2.42]$, while the 95\% confidence interval based on two-stage least squares is $[-0.94, -1.94]$. For presidential elections from 1980 to 2000, the confidence interval for the almost exact method is noticeably wider, in fact the almost exact interval, $[-1.26, -5.19]$ is almost twice as long as the interval from TSLS, $[-1.18, -3.25]$, and the Bloom method, $[-1.59,-3.25]$. Finally, we restrict the data to the period from 1984 to 2000, the differences in confidence intervals are quite stark. Now the almost exact method returns an interval that covers from $-\infty$ to $\infty$. The intervals from TSLS and the Bloom method are wider than before, but both are closed intervals. For example the TSLS 95\% confidence interval is $[-0.037, -5.99]$, and the 95\% confidence interval based on the Bloom method is $[-3.27, -5.99]$. The confidence interval from the almost exact approximation provides a clear warning that the instrument in this case is weak. It is worth noting that in this analysis there are nearly 10,000 observations, so the sample size is more than adequate. The lack of statistical certainty, here, is driven almost entirely by the weakness of the instrument.
    
\begin{table}
\centering
\caption{Point estimates and confidence intervals for analysis of rainfall as an instrument for voter turnout.}
\label{tab:rain}
\begin{tabular}{lcccccc}
\hline
 & Elections 1976--2000 & Elections from 1980--2000 & Elections from 1984--2000  \\ 
\hline
N & 13687 & 11729 & 9770 \\ 
Point Estimate & -1.4 & -2.2 & -4.6  \\ 
Almost Exact 95\% CI & [-0.91, -2.42]  & [-1.26, -5.19] & [-$\infty$, $\infty$]  \\ 
TSLS 95\% CI & [-0.94, -1.94]  & [-1.18, -3.25] & [-0.037, -5.99]  \\  
Bloom 95\% CI & [-1.04, -1.85] & [-1.59, -3.25] & [-3.27, -5.99] \\   
\hline
\end{tabular}
\end{table}

\section{Discussion}

In this paper, we have reviewed inferential methods for the instrumental variables method, with a focus on exact methods. In particular, we highlighted exact and almost exact methods, which see little use in practice, but have important advantages. We used a Monte Carlo study to show that the almost exact method maintains the nominal coverage rate even when the instrument is quite weak. In contrast, methods based on Normal approximations had poor coverage in the same setting. However, when the instrument is strong and sample sizes are large, all the methods provide very similar results, both in simulations and empirical applications. In fact, the Normal approximation via the Bloom method seems to provide the simplest form of inference for $\tau$ under this case.  The Bloom method still sees widespread use when analysts attempt to convey results to nontechnical audiences. This appears to be a safe practice when applied to a randomized encouragement design with one-sided noncompliance rates that do not fall below 25\%.  While we do not provide systematic evidence on this point, we suspect that in most randomized policy interventions compliance rates are typically not this low.  However, in observational studies, the likelihood of weak instruments is greater and exact or almost exact methods should see more use by applied analysts.

One additional method of inference that may be applied to IV estimators is the bootstrap. We did not consider the bootstrap in this article, since we confined ourselves to finite population inference, and the bootstrap typically assumes an infinite population model.  Moreover, the smoothness condition required for the bootstrap may fail when the instrument is weak. Finally, while the bootstrap is generally second order accurate, that is not always the case for IVs \citep{horowitz2001bootstrap}.  In our opinion, unless investigators are interested in population inferences, exact or almost exact methods are generally preferred over the bootstrap when applied to IV estimators.

One area of applied study where the problem of weak instruments is common is in the study of genetics. In studies of Mendelian randomization (MR), the IV method has become a standard analytic tool. In these studies, the source of the instruments are genetic variations, and the compliance rates, or in MR context, the explained genetic variations, can be very low. While there has been work on weak instruments within the MR context \citep{burgess_bias_2011,burgess_avoiding_2011,pierce_power_2011}, we believe our work here, specifically the almost exact method with its attractive formula and robustness guarantees, can complement some of the proposals to deal with the problem of weak instruments in MR.

We provide an \texttt{R} function in the Supplementary Materials that returns confidence intervals under the almost exact method for use by applied researchers. One additional advantage of almost exact methods is that they can be combined with rank based test statistics when the outcome distribution is heavy tailed or have unusual observations \citep{Rosenbaum:1996}. When a rank-based test statistic is used, asymptotic approximations to the randomization distribution again provide convenient results when sample sizes are larger.

\singlespacing
\bibliographystyle{asa}
\bibliography{fixed_iv}
\end{document}